\documentclass[%
 reprint,
nofootinbib,
 amsmath,amssymb,
 aps,
]{revtex4-2}
\usepackage{ads}
\usepackage{float}
\usepackage{graphicx}% Include figure files
\usepackage{dcolumn}% Align table columns on decimal point
\usepackage{bm}% bold math

\usepackage{color}

\begin{document}

\preprint{APS/123-QED}

\title{Improvement of cosmological constraints with the cross correlation between line-of-sight optical galaxy and FRB dispersion measure}% Force line breaks with \\

\author{Chenghao Zhu}
\affiliation{Physics Department, Fudan University, Shanghai 200433, China}%Lines break automatically or can be forced with \\
\author{Jiajun Zhang}
\email{jjzhang@shao.ac.cn}
\affiliation{Shanghai Astronomical Observatory, Chinese Academy of Sciences, Shanghai 200030, China}
\date{\today}% It is always \today, today,

\begin{abstract}
Fast Radio Bursts (hereafter FRBs) can be used in cosmology by studying the Dispersion Measure (hereafter DM) as a function of redshift. The large scale structure of matter distribution is regarded as a major error budget for such application. Using optical galaxy and dispersion measure mocks built from N-body simulations, we have shown that the galaxy number density can be used as a tracer for large scale electron density and help improve the measurement of DM as a function of redshift. We have shown that, using the line-of-sight galaxy number counts within 1' around the given localized FRB source can help improve the cosmological parameter constraints by more than 20\%.
\end{abstract}

\maketitle

%\tableofcontents

\section{\label{sec 1}INTRODUCTION }
Currently, $\Lambda$CDM model is considered as the standard model of cosmology \cite{bull2016beyond}. It is supported by varieties of observations, including cosmic microwave background temperature anisotropy measurement from Planck \cite{collaboration2020planck}, baryon acoustic oscillations data from SDSS galaxy surveys \cite{ross2015clustering}, type Ia supernovae data from the Joint Light-curve Analysis \cite{betoule2014improved}, and $H_0$ data from distance ladder observation \cite{freedman2019carnegie,macri2006new}. The constrains of cosmological parameters are quite precise today, but still lots of attempts are made to verify the consistency from different methods. There is some tension in the measurement of $H_0$ \cite{Riess2021ApJH0,Valentino2021H0,perivolaropoulos2021challenges,elcio2022H0}. Therefore, we demand new methods probing the cosmic expansion history, such as using standard siren, 21cm intensity mapping, FRBs \cite{jin2021PhRvD,qiu2022JCAP,wang2022SCPMA,wu2022cosmo}.

Fast Radio bursts are short-duration ($\sim$ms) radio bursts ($\sim$GHz) which was first detected in 2007 \cite{lorimer2007bright}. FRBs seems to be very common in our universe \cite{fialkov2017fast}. From their isotropic dispersion measures, most FRBs have been confirmed to have extragalactic origins \cite{xu2015extragalactic}. FRB200428 is the first FRB which is located inside the Milky Way \cite{andersen2020bright}. The astrophysical mechanism of FRBs is still mysterious, but more and more information about their source have been reported \cite{spitler2016repeating,andersen2020bright}. Several good reviews have summarized current theories about FRBs and its physical mechanism \cite{xiao2022fast,platts2019living,zhang2020physical}.  Moreover, hundreds of FRBs are gradually observed nowadays. Their information are available online \footnote{\url{http://www.frbcat.org}}. Many radio telescopes such as CHIME \cite{amiri2021first}, PARKES \cite{price2018breakthrough}, FAST \cite{niu2021crafts}, BINGO \cite{abdalla2021bingo}, SKA \cite{skafrb2020MNRAS} have detected considerable FRBs or can be expected to observe much more FRBs in the near future. %ska fast bingo tianlai chime parkes

Since the first known repeater FRB121102 \cite{spitler2016repeating} was detected and located, with or without a repeating source, several FRBs are well located as well \cite{bannister2019single,ravi2019fast}. With their extragalactic location known, FRBs show great potential to be new independent probes of cosmology \cite{munoz2018finding,prochaska2019probing}. The dispersion measure (hereafter DM) of an FRB contains the information about ionized baryons of the path from the source to earth.  The expectation of DM contributed by the intergalactic medium from an FRB at redshift z is highly related to the cosmological model and parameters. Hence the FRBs with their DM(z) can be used to contrain cosmology, although it is still hard to separate DM from its host galaxy's contribution.

Previously, \citet{macquart2020census} has tried to do parameters constrain from known localized FRBs in the universe. Combined with constrains from other observations \cite{collaboration2020planck} as their prior, \citet{walters2018future} explore the constrains with more FRBs generated by simulation. Futher, \citet{walters2019probing} discussed the result with a weaker assumption of knowledge about diffuse gas fraction.

From their large dispersion measure, naturally we can associate FRBs with large scale structure. \citet{rafiei2021chime} has summarized the angular cross-correlation between DM and large-scale structure of galaxy with the results released by CHIME/FRB catalogue \cite{amiri2021first}. In addition, \citet{shirasaki2022probing}' investigated the angular cross-correlation in detail with hydrodynamical simulation illustrisTNG. Following their inspiration, we attempted to use such a cross-correlation to partly eliminated the variance of DM contributed by the large-scale structure, in order to obtain a better cosmological constraints. In this work, we will verify the improvement of cosmological parameter constraints using FRBs with the knowledge of line-of-sight galaxy number.

In this paper, we will show our results as follows:
in section \ref{sec 2}, we will introduce some basic background knowledge, including the definition of DM and how to calculate it in different coordinate; in section \ref{sec 3}, we will introduce the method how we deal with the N-body simulation data; in section \ref{sec 4}, we will show some noteworthy results from mocks; in section \ref{sec 5}, we will show the concrete MCMC improvement after applying the cross-correlation we acquire; finally in section \ref{sec 6}, a brief summary and discussion will be given.

\section{\label{sec 2}PRELIMINARIES}

\begin{figure}
    \centering
    \includegraphics[width=1.0\columnwidth]{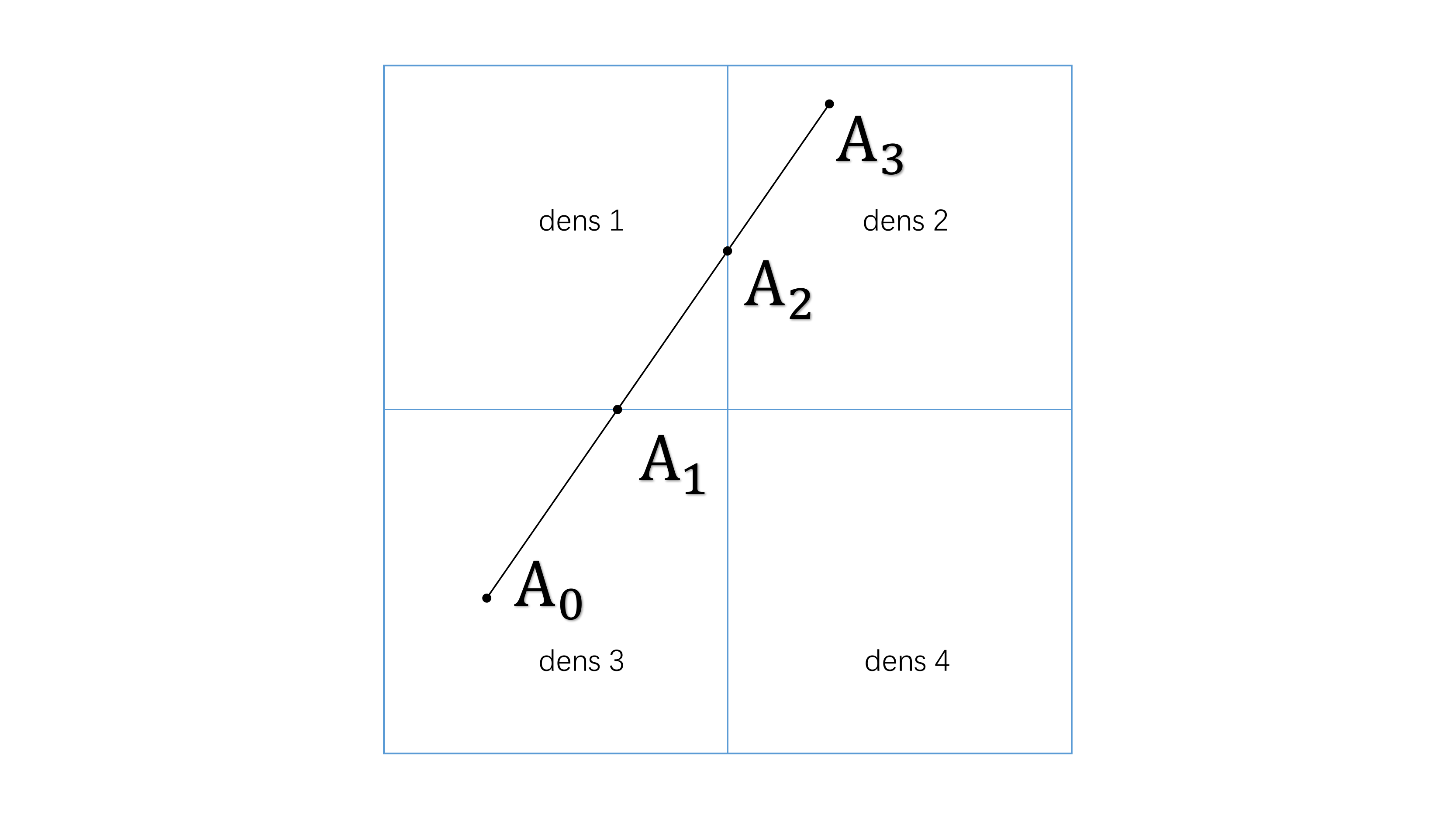}
    \caption{Without loss of generality,  we display a 2-dimensional case for its convenience to illustrate. Here we set $A_0$ to be our 'observer' and $A_3$ to be the FRBs source. dens i (i=1,2,3,4) are the density given by MDR1 database, $A_1$ and $A_2$ are intersection points of path with the grids.}
    \label{fig method}
\end{figure}
In general, DM represents the delay time of a signal traveling through the plasma, which is determined by the dispersion effect of the plasma:

\begin{eqnarray}
\Delta t &&= \frac{e^2}{2\pi m_ec}(\frac{1}{v^2_1}-\frac{1}{v^2_1})\int \frac{n_e}{1+z} \, dz\nonumber\\
&&\simeq 4.15s[(\frac{\nu_1}{1GHz})^{-2}-(\frac{\nu_2}{1GHz})^{-2}] \frac{DM}{10^3}pc\cdot cm^{-3}.
\label{eq defDM}
\end{eqnarray}

It is noteworthy that, in the rest frame, DM is defined as $DM'=\int n_e(z)\, dl$, which is the column density of free electrons along the path. But in the observer frame, the observed delay time is $\Delta t=\Delta t' \times (1+z)$ and the observed frequency is $\nu = \nu'/(1+z) $. Consequently, the measured DM by an earth observer is modified as $DM=\int \frac{n_e(z)}{1+z}\, dl$. \cite{deng2014cosmological}

The total DM of an FRB can be separated as 
\begin{equation}
    DM=DM_{MW}+DM_{IGM}+\frac{DM_{HG}}{1+z}
    \label{eq contri}
\end{equation}

where we denote Milky Way, intergalactic medium and host galaxy by MW, IGM and HG. In actual observation, contribution of Milky Way can be obtained from research about pulsars, though not very accurately. Usually we cannot get adequate information of host galaxy of an FRB, and the contribution of DM from the host galaxy is very likely unknown. While the contribution of IGM, which is highly associated with large scale structure, can provide information about cosmology.

\subsection{Dispersion Measure in Comoving Coordinates}
We will use N-body simulation data to generate mocks of DM distribution in this study. Since the simulation has been performed in comoving coordinates, we need to obtain expression of DM in comoving space. Notice that the unit of $n_e$ is $1/pc^3$ and the unit of $dl$ is cm. Thus at redshift z, $n_e(z)=(1+z)^3n_e(0)$, $dl_{z=z'}=dl_{z=0}/1+z'$. Then we have
\begin{equation}
    DM=\int_0^\chi n_e(\chi ')(1+z(\chi ')\,d\chi ',
    \label{eq DMcomov}
\end{equation}
where we used the relation between comoving distance and physical distance $d\chi =dl_{z=0}$.

\subsection{Mean Dispersion Measure from IGM}
The mean DM contributed by IGM can simply derived from its definition. In this article, we apply $\Lambda$CDM cosmology. Then, in the comoving ordinate, the average number density of electron at redshift z can be expressed as
\begin{equation}
    \bar{n}_e=\rho_{c,0} (1+z)^3 \Omega_b \times (\frac{3}{4} \frac{I_H(z)}{m_p}+\frac{1}{4} \frac{I_{He}(z)}{4m_p}),
    \label{eq nmean}
\end{equation}
where $\rho_{c,0}=\frac{3 H_0^2}{8\pi G}$ is the critical density of the universe today, $\Omega_b$ is fraction of baryon, $I_H(z)$ is the average number of ionized electron in the IGM emitted by one hydrogen atom, and $I_{He}$ denotes the contribution of helium.

The relation between physical distance interval and redshift z is given by
\begin{eqnarray}
dl&=& \frac{1}{1+z} c H(z) dz\nonumber\\
&=&\frac{1}{1+z} \frac{c}{H_0} \frac{dz}{ \sqrt{ (1+z)^3\Omega_M+\Omega_\Lambda } }.
\label{eq dl}
\end{eqnarray}
Here we use $\Lambda$CDM model again and assumed flat universe, thus $\Omega_\Lambda=1-\Omega_M$. Plugging back into the definition of DM, we finally obtain
\begin{equation}
    \overline{DM}_{IGM}=\frac{3c\Omega_b H_0}{8\pi Gm_p}\int_0^z(1+z')\frac{I(z')}{E(z')}\,dz',
    \label{eq DMIGM}
\end{equation}

where $I(z')=\frac{3}{4} I_H(z)+\frac{1}{16}I_{He}(z)$ and $E(z')=\sqrt{(1+z)^3\Omega_M+\Omega_\Lambda}$.

\section{\label{sec 3} Method}
\begin{figure*}[!t]
\begin{tabular}{cccc}
    DM &
    \begin{minipage}{0.6\columnwidth}
    \centering
    \includegraphics[width=1\columnwidth]{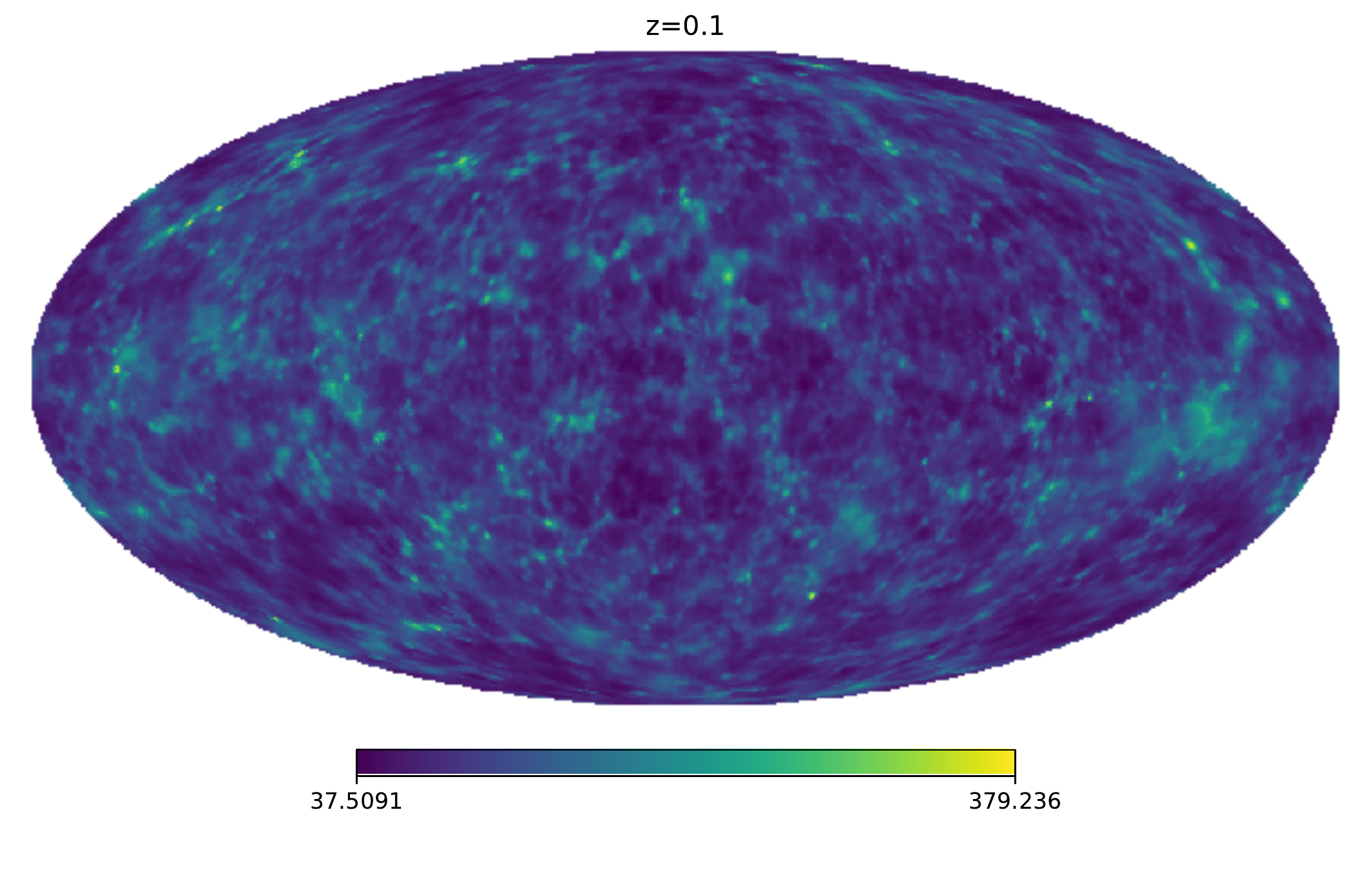}
    \end{minipage}&
    \begin{minipage}{0.6\columnwidth}
    \centering
    \includegraphics[width=1\columnwidth]{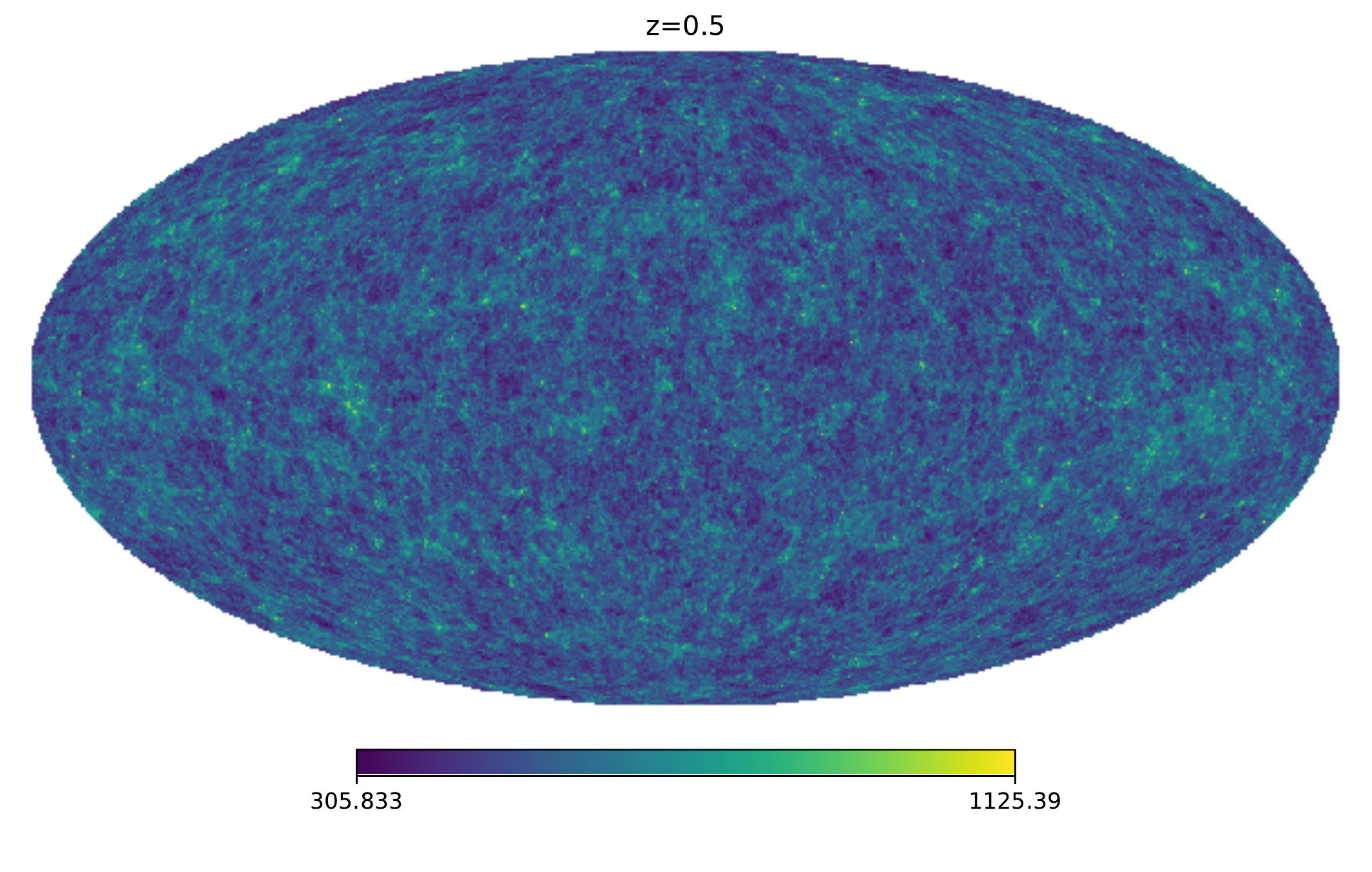}
    \end{minipage}&
    \begin{minipage}{0.6\columnwidth}
    \centering
    \includegraphics[width=1\columnwidth]{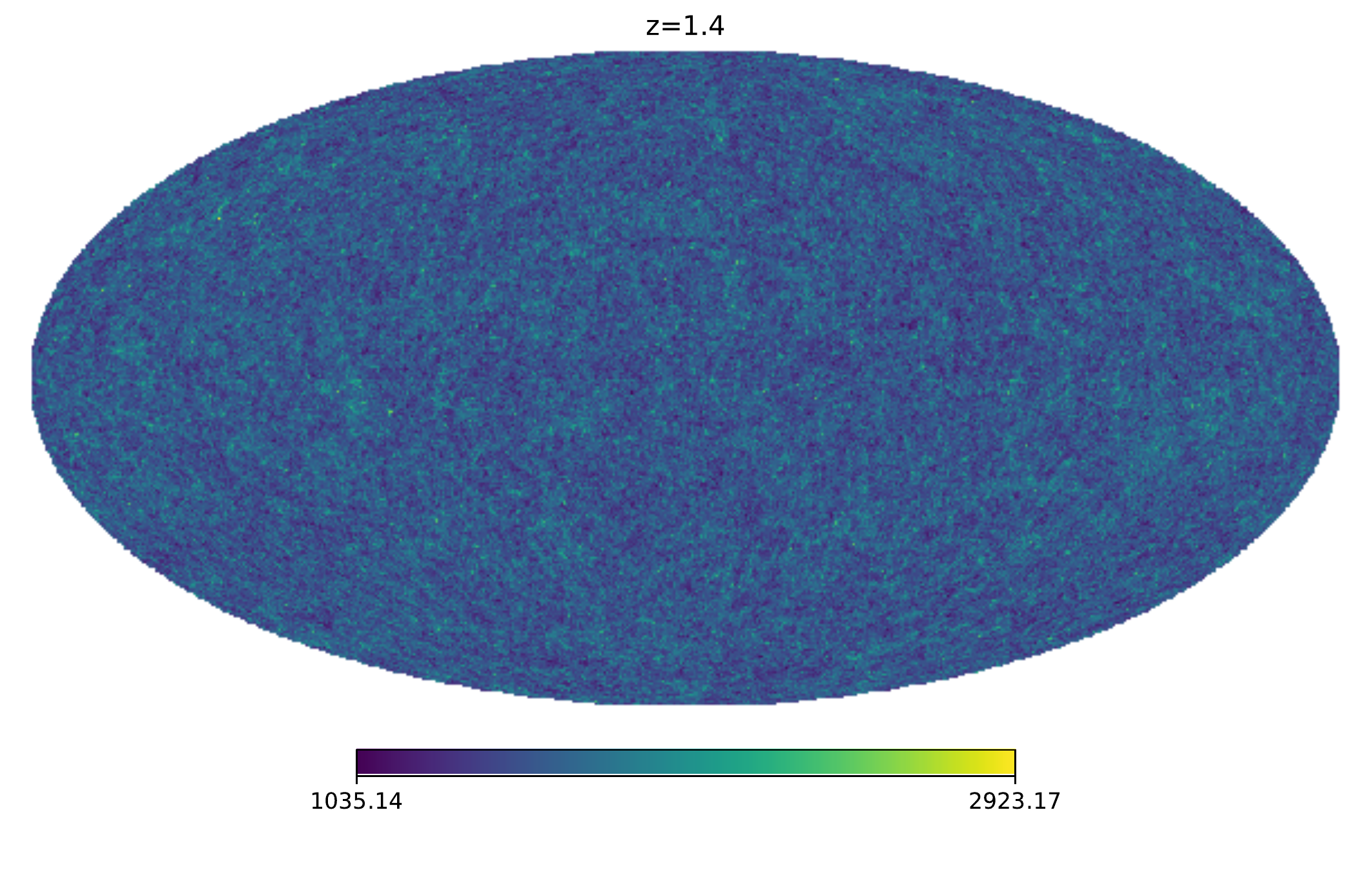}
    \end{minipage}\\
    cone&
    \begin{minipage}{0.6\columnwidth}
    \centering
    \includegraphics[width=1\columnwidth]{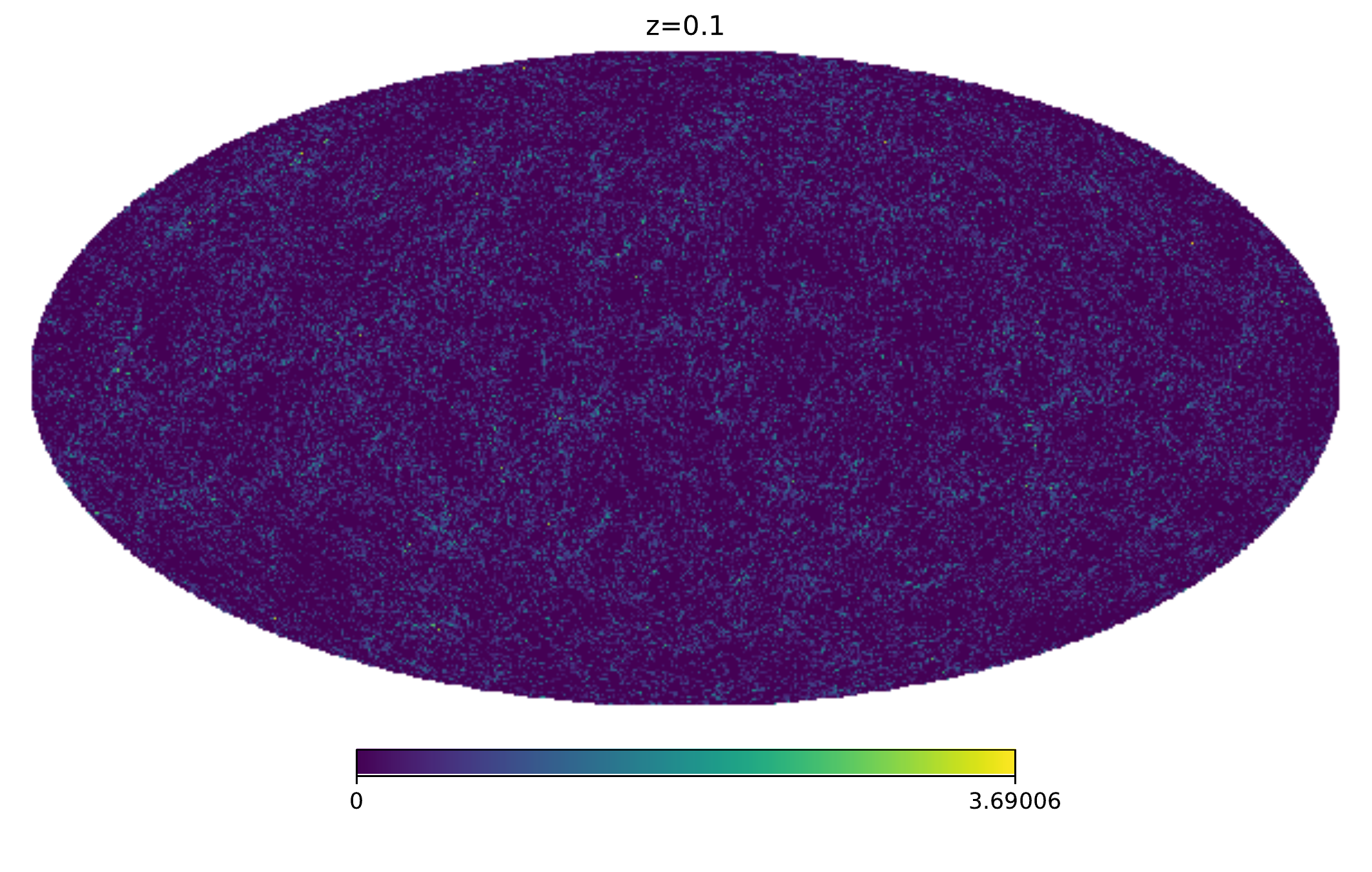}
    \end{minipage}&
    \begin{minipage}{0.6\columnwidth}
    \centering
    \includegraphics[width=1\columnwidth]{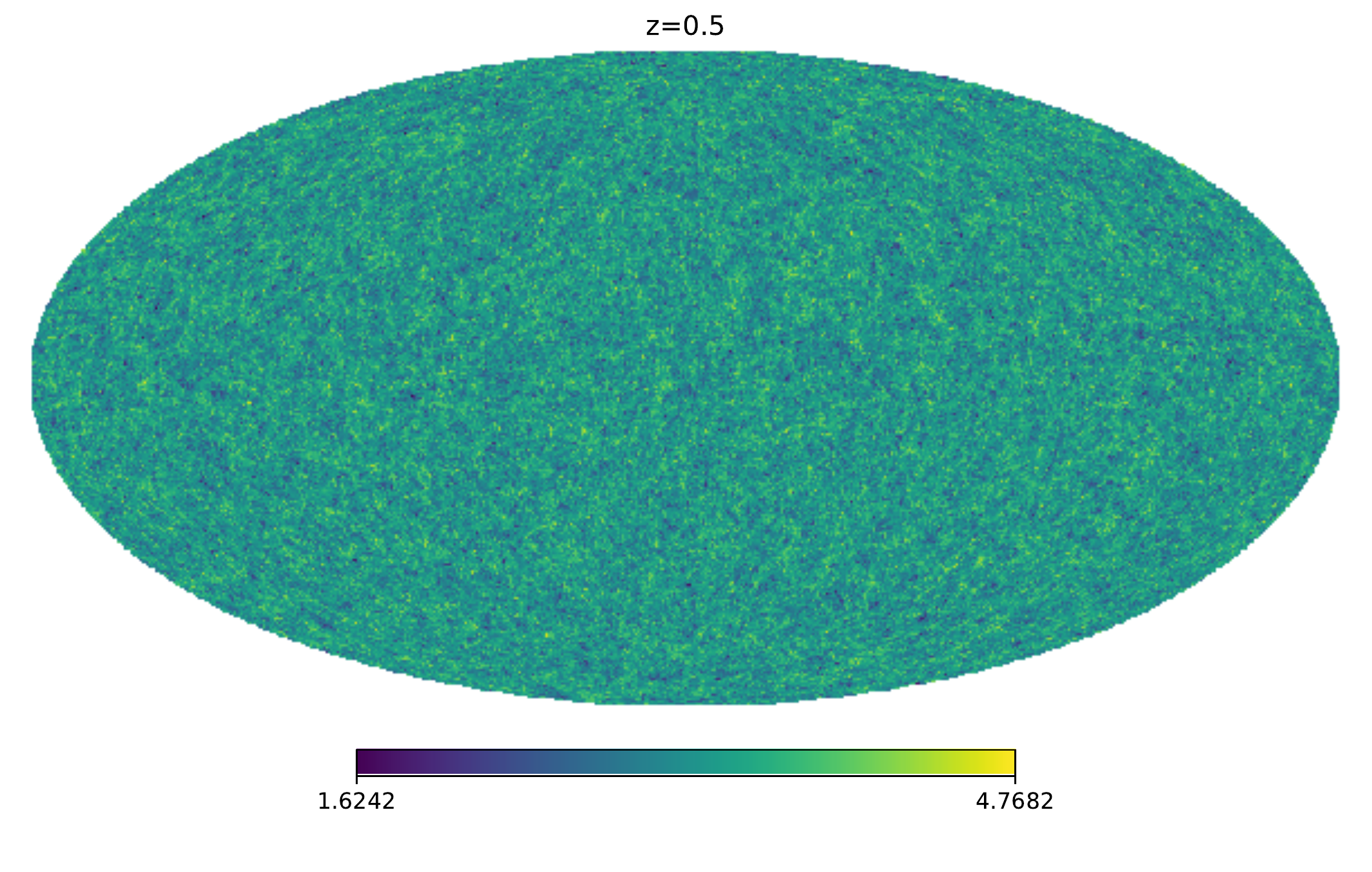}
    \end{minipage}&
    \begin{minipage}{0.6\columnwidth}
    \centering
    \includegraphics[width=1\columnwidth]{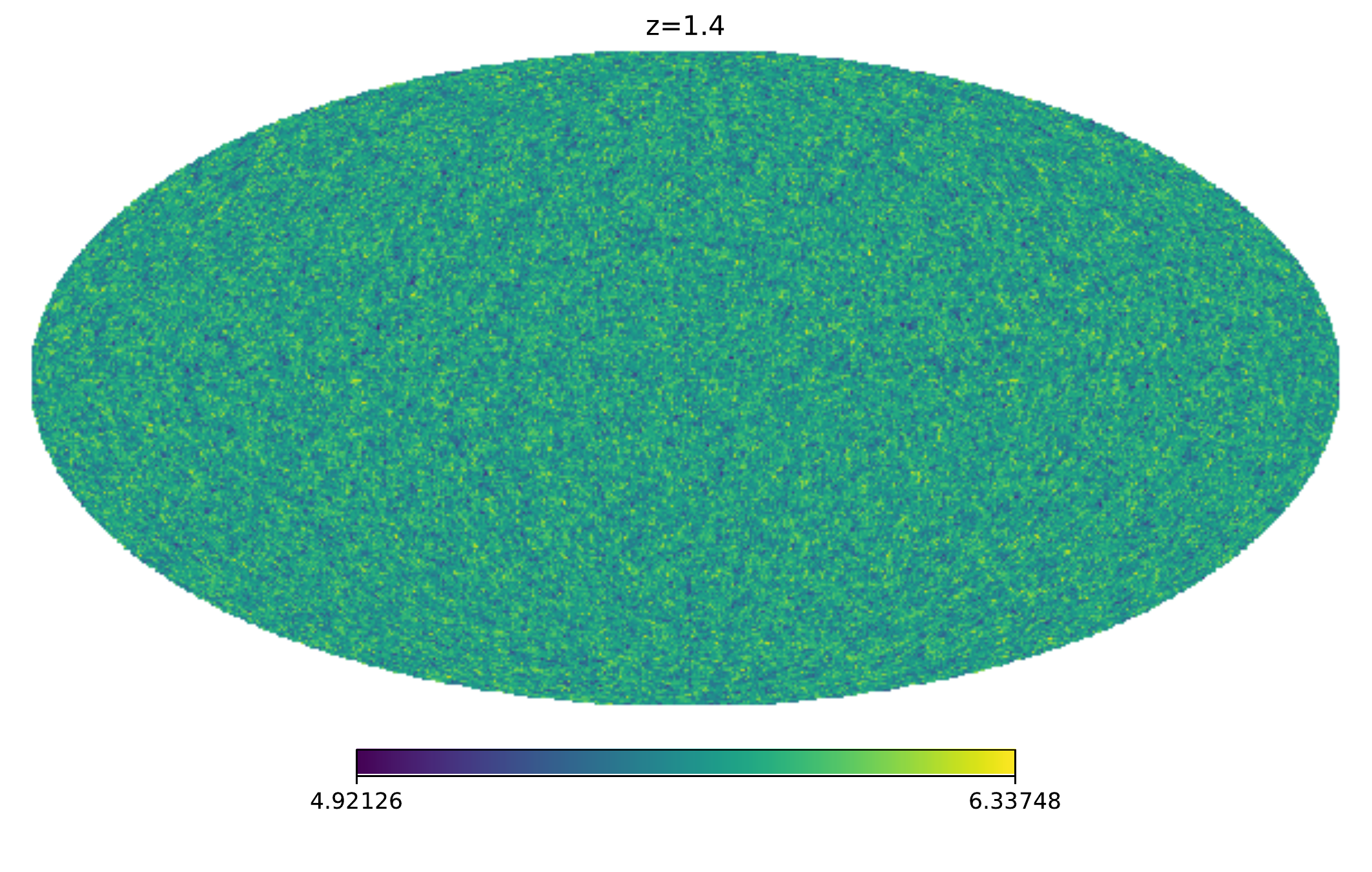}
    \end{minipage}\\
    cone+&
    \begin{minipage}{0.6\columnwidth}
    \centering
    \includegraphics[width=1\columnwidth]{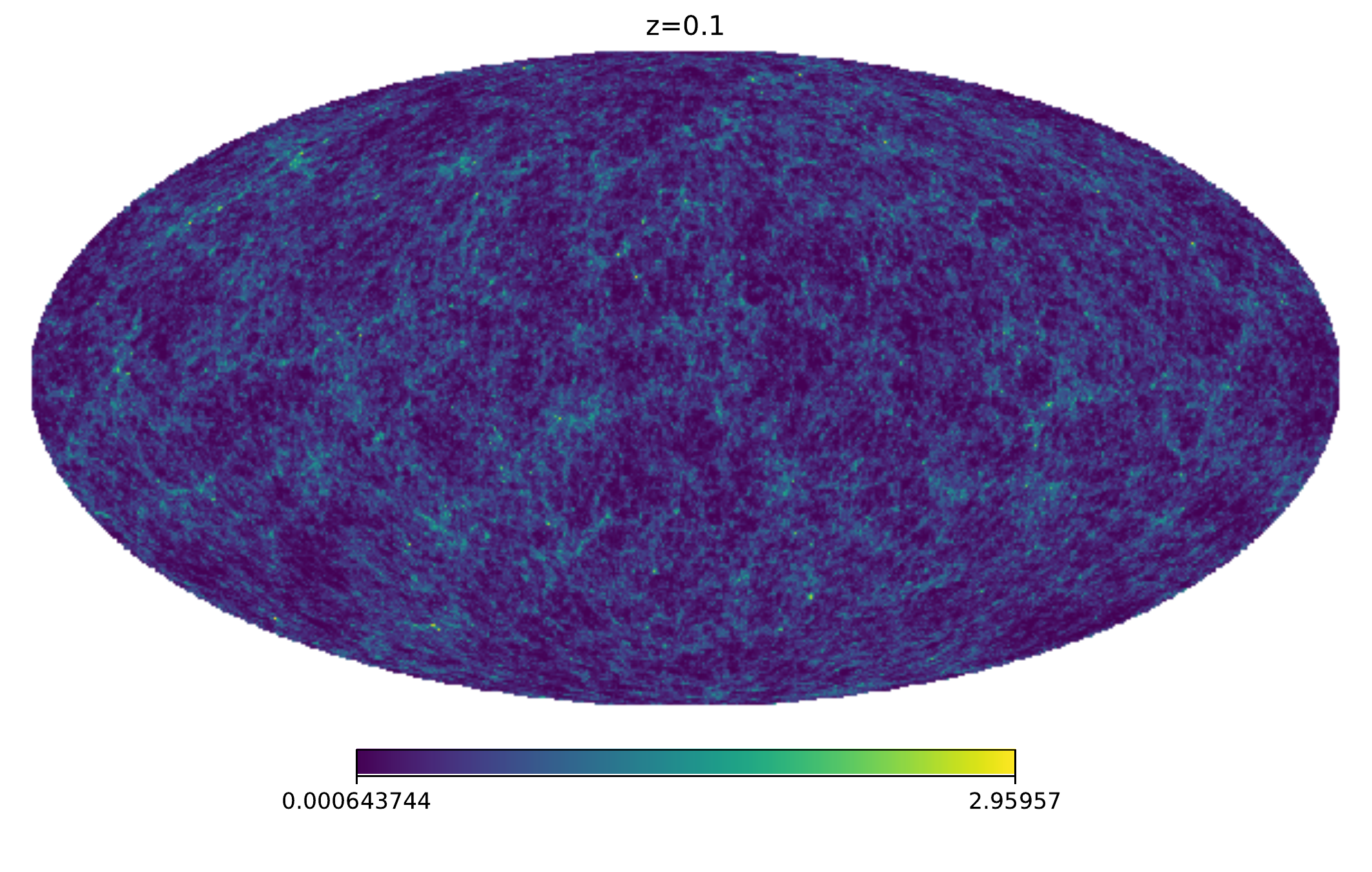}
    \end{minipage}&
    \begin{minipage}{0.6\columnwidth}
    \centering
    \includegraphics[width=1\columnwidth]{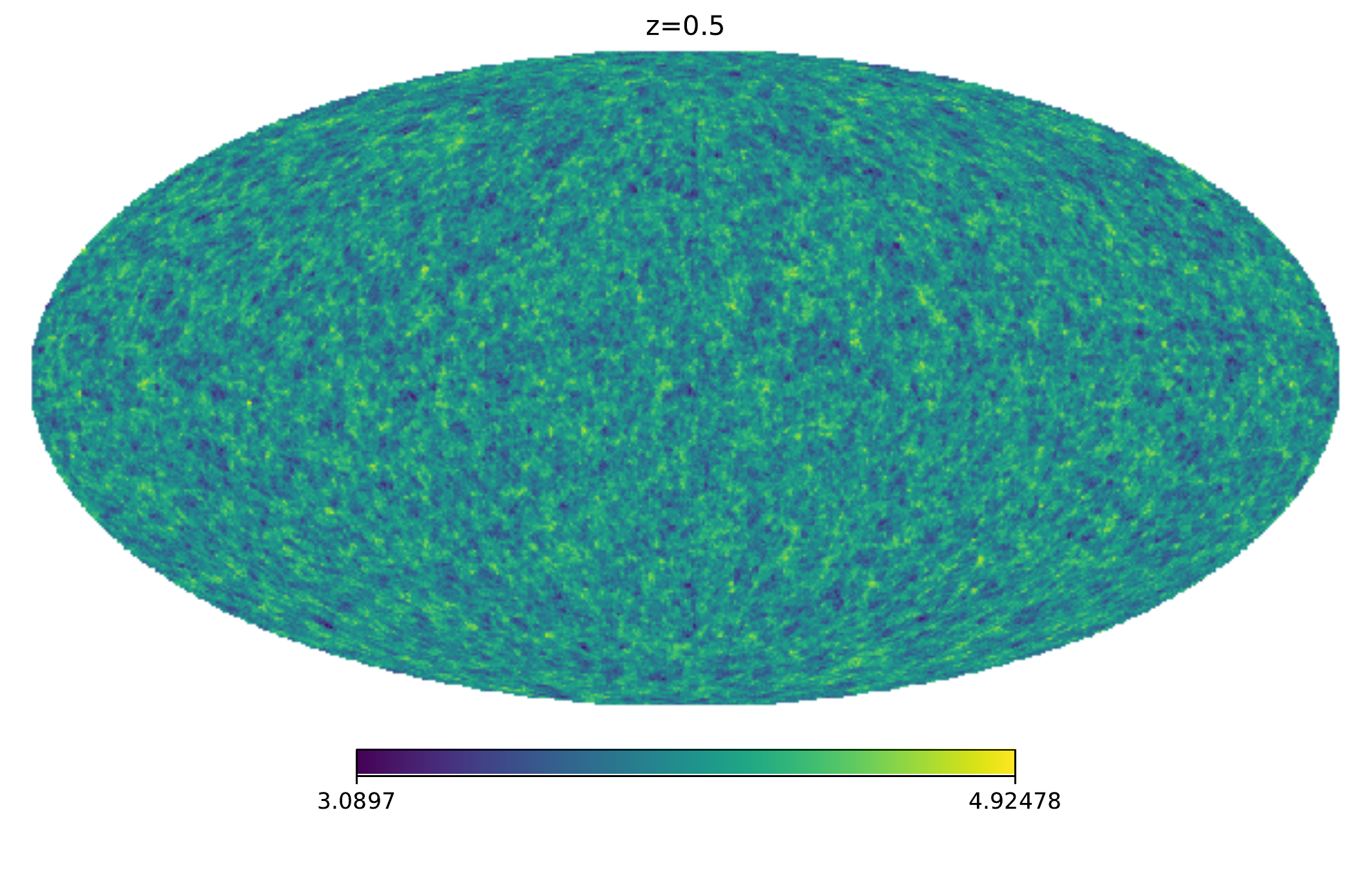}
    \end{minipage}&
    \begin{minipage}{0.6\columnwidth}
    \centering
    \includegraphics[width=1\columnwidth]{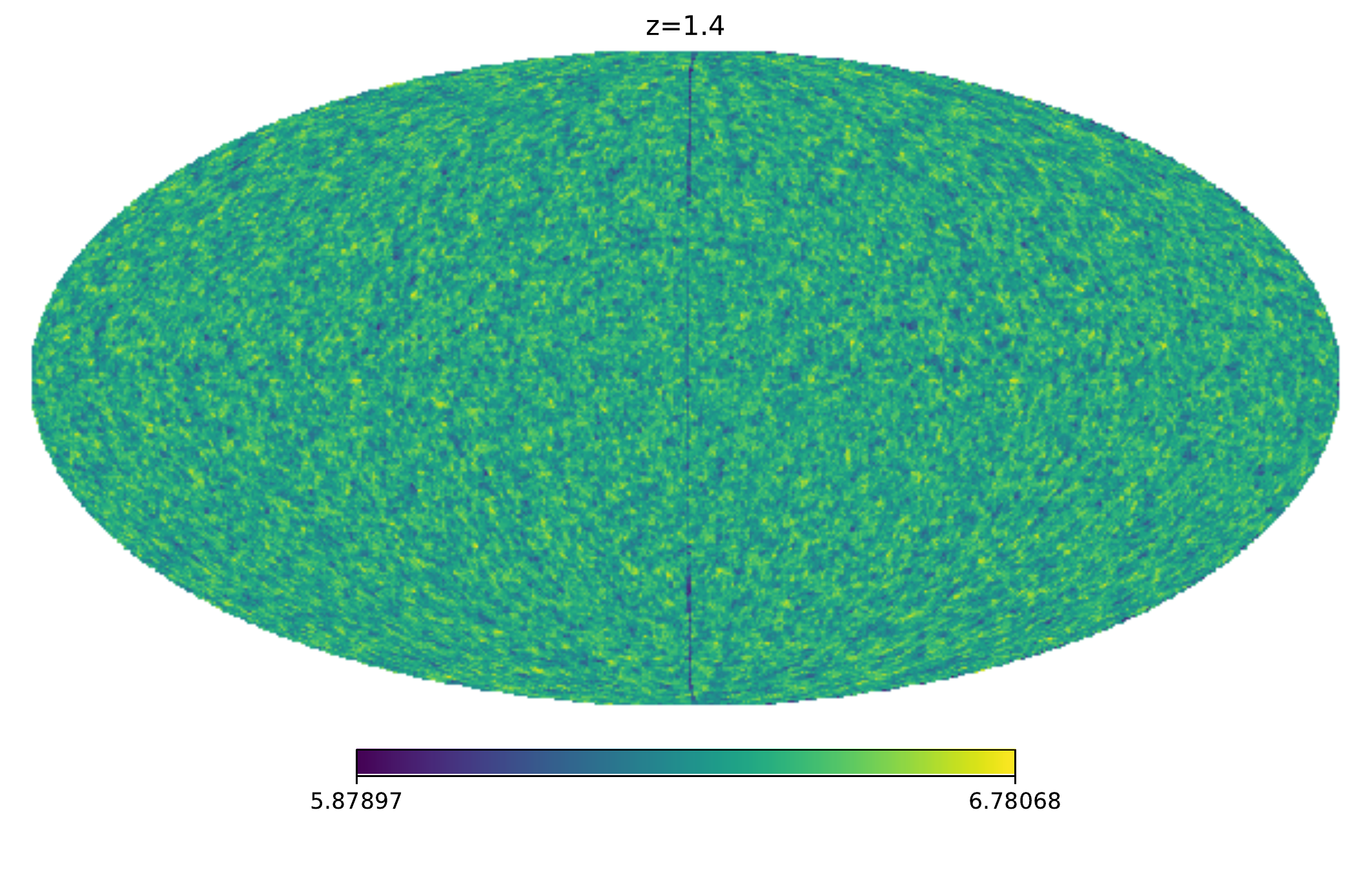}
    \end{minipage}
\end{tabular}
\caption{From left to right, we plot the mollweide projection images from the observer to z=0.1, 0.5 and 1.4. From up to bottom, we plot the mollweide projection images of DM, galaxy number with "cone" method and galaxy number with "cone+" method. Notice that we take logarithm for illustrating $n_{gal}$, i.e. the value of each pixel is $\ln (1+n_{gal})$ for both 'cone' and 'cone+'. }
\label{}
\end{figure*}

MDR1 is the first MultiDark simulation, performed in 2010\cite{francisco2012halo}. This simulation contains about 8.6 billion particles in a $(1\, Gpc/h)^3$ cube, using WMAP5 cosmology. MDR1 provides the matter density field data and the halo catalogue. In our research, we need to calculate the dispersion measure between arbitrary two points in the box and  the galaxy number in an arbitrary light-cone (or weighted light-cone which we will introduce later).

The side length of the cosmological cube is 1 Gpc/h, which is still too small to cover the redshift range of FRBs. We have used the periodic boudary condition of the simulation to construct a much larger box to cover all the FRBs we have considered, up to $z=2$. 

MDR1 only gives the data of over-density in each cell (for convenience we will use density instead, and denote it by $dens=\rho_{background}\times(1+overdensity)$). To calculate DM, we need to obtain number density of free electron in each cell. For simplicity, we have set the distribution of baryon to be the same as dark matter. We ignored the difference in metallicity and ionization fraction, assumed that $I_H=1$ and $I_{He}=2$. Then expression of DM becomes
\begin{equation}
    DM=\int_0^\chi dens(\chi' ) \times \frac{7}{8}\frac{\Omega_b}{\Omega_m}\,d\chi'.
    \label{eq DMMDR1}
\end{equation}

As shown in Fig.\ref{fig method}, we have illustrated how to calculated DM between arbitrary two points from the density fields, 
\begin{eqnarray}
DM(\overline{A_0A_3})&=&\int_{A_{0}}^{A_{3}} n_e(\chi ')(1+z)\,d\chi '\nonumber\\
&=&dens_3 \int_{A_{0}}^{A_{1}}(1+z)\,d\chi' +dens_1 \int_{A_{1}}^{A_{2}}(1+z)\,d\chi'\nonumber \\
&+&dens_2 \int_{A_{2}}^{A_{3}}(1+z)\,d\chi'.
\label{eq DMmethod}
\end{eqnarray}

We use halo occupation distribution\cite{zheng2005theoretical} to transform halo catalogue into galaxy number. We choose Favole's model\cite{favole2015building} to obtain galaxy number of each halo,
\begin{eqnarray}
    N_{cen}(M)&=&\frac{1}{2}[1+erf(\frac{\log M-\log M_{min}}{\sigma_{\log M}})]\nonumber\\
    N_{sat}(M)&=&N_{cen}(M)(\frac{M-M_0}{M_1'})^\alpha,
    \label{eq HOD}
\end{eqnarray}
where $erf(x)=2\int_0^xe^{-t^2}/\sqrt{\pi}\,dt$ is the error function. In our research, we set up 5 parameters of Halo occupation distribution as follows: $M_{min}=10^{12.681}, M_0=10^{12.296}, M_1=10^{13.635}, \sigma_{M}=0.532$ and $\alpha=0.994$. Then the average galaxy number is $\bar{n}_{gal}=0.0012/(Mpc/h)^3$.

\section{\label{sec 4}Mock Data}

Since in the simulation, DM can be directly calculated, we randomly choose a point in the box to be our observer, and obtain the DM from all angles of the full sky to different redshift. Following the method illustrated above, we get 20 global maps of DM in total from $z=0.1$ to $z=2$ equidistantly with corresponding maps of galaxy number (hereafter $n_{gal}$). With these data, we can give a detailed discussion about the correlation to prepare for further analysis. 
\subsection{Line-of-sight Galaxy Number}
We use two different methods to obtain $n_{gal}$. In the first way which we denote by 'cone', we sum those galaxies whose halo center is contained in our light-cone with half top angle $\theta$. In the second way denoted by 'cone+', we count all galaxies whose position vector's projection on the axis pointing from observer to source is positive, but with a factor exponentially decay about its perpendicular distance to the axis, that is, $exp[-(||\mathbf{v}||\sin \theta)/(\mathbf{v}_z \tan \theta)]$, where $\mathbf{v}$ is the position vector, $\mathbf{v}_z$ is its projection to the axis and theta is the half top angle of the 'cone+'.

In Fig.\ref{fig mollview}, we have shown the Mollweide projection map of DM, "cone" galaxy number and "cone+" galaxy number from the observer to $z=0.1, 0.5$ and $1.4$. We can clearly see that the DM map strongly correlate with the galaxy number distribution, which is what we expected. At high redshift such as $z=1.4$, it is hard to identify correlation between DM and galaxy number from the map. We will show quantitative results of cross correlation between DM and galaxy number in the next subsection.
\subsection{Cross Correlation}

\begin{figure}
    \centering
    \includegraphics[width=1\columnwidth]{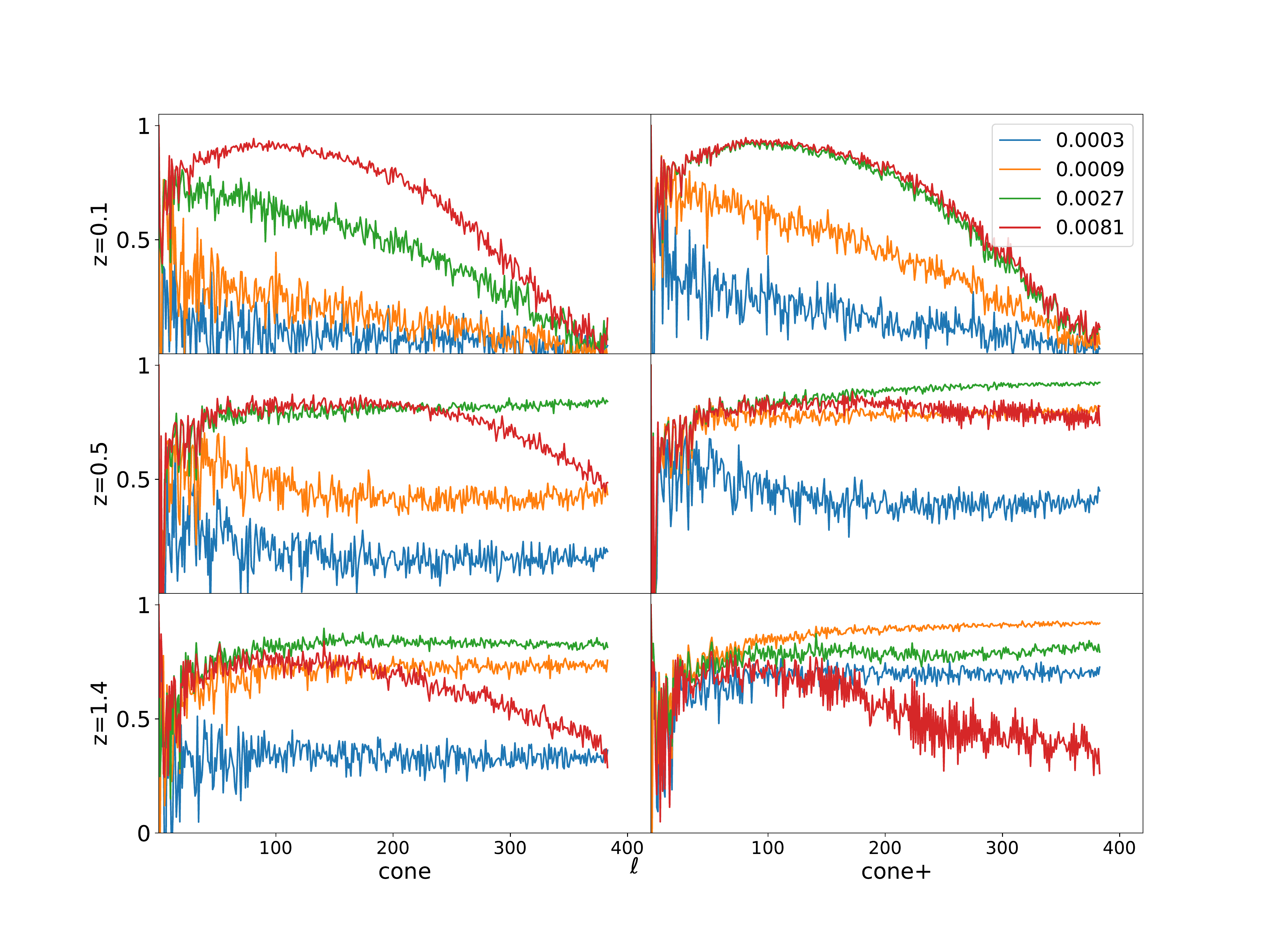}
    \caption{The cross-correlation coefficient of both galaxy number calculating method at three different redshifts. Different colors represent different line-of-sight angle $\theta$ for the counting threshold. We have verified the influence of angle (with unit radiant) at redshift z=0.1, 0.5, 1.4. For both 'cone' and 'cone+', when the half top angle $\theta$ is 0.0027 (radiant), a good cross-correlation holds.}
    \label{fig difftheta}
\end{figure}

\begin{figure}
    \begin{minipage}{1\columnwidth}
    \centering
    \includegraphics[width=1\columnwidth]{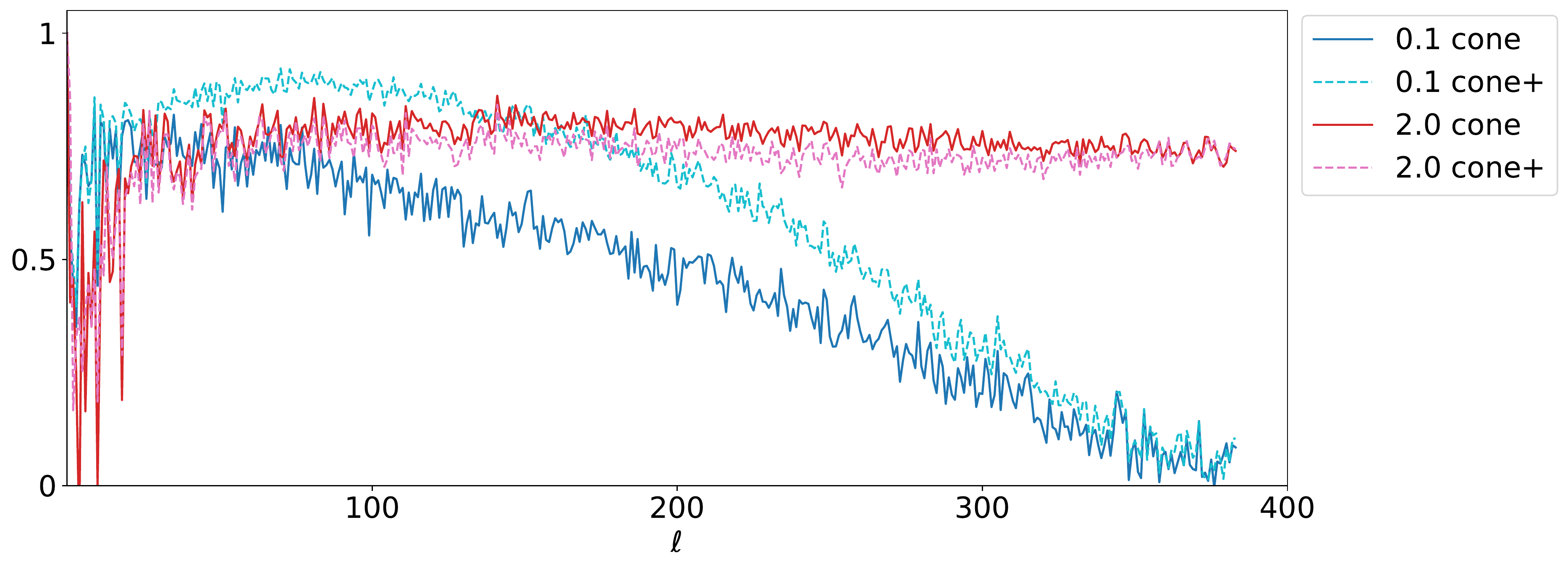}
    \end{minipage}
    \begin{minipage}{1\columnwidth}
    \centering
    \includegraphics[width=1\columnwidth]{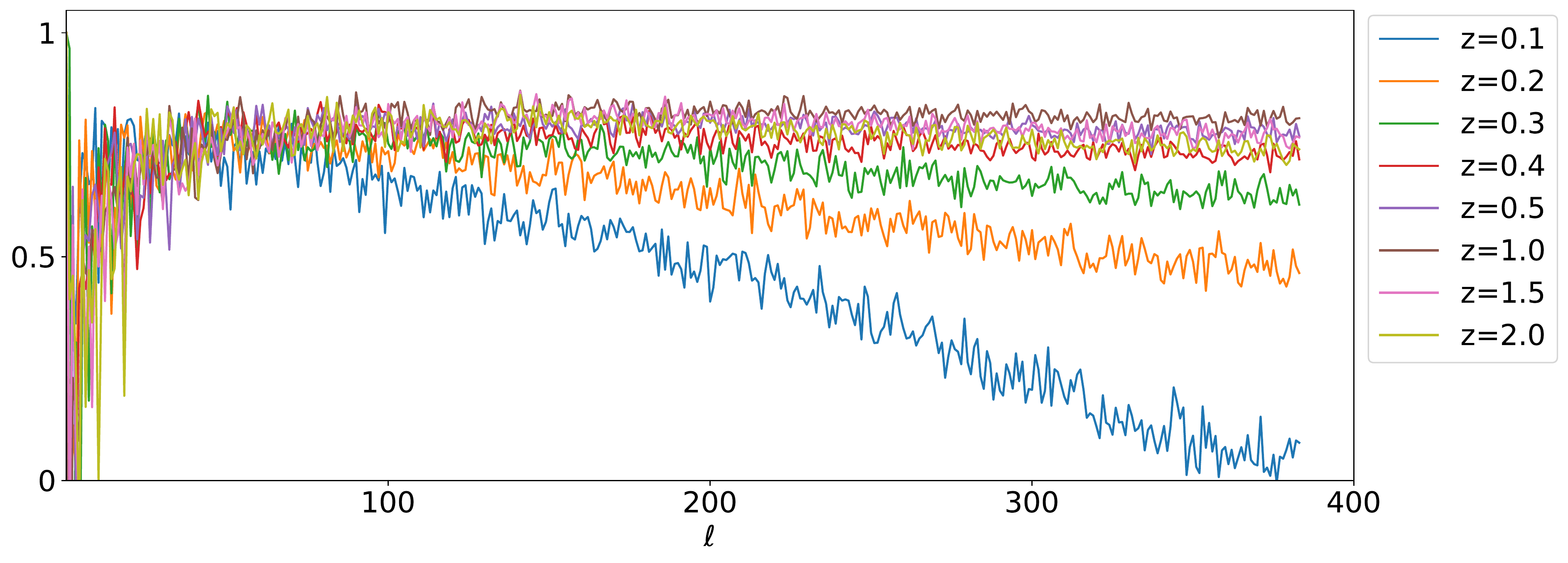}
    \end{minipage}
    \caption{In the top panel, we show the difference between 'cone' and 'cone+'. At low redshift, using 'cone+' can obtain a better correlation, but at high redshift this difference is very small. In the bottom panel, we show the redshift evolution of the cross correlation. The coefficient monotonically increase with the redshift when the redshift is low, and become stable after z=0.5.}
    \label{fig conevscone+}
\end{figure}

The value of $n_{gal}$ of each pixel depend on the ling-of-sight angle $\theta$ we set, and so does the correlation between DM and $n_{gal}$. Hence we have tried with different angles at red-shift z=0.1, 0.5, 1.4, which we think can reflect the behavior in the whole range we want to investigate.

The cross correlation coefficient between DM and $n_{gal}$ for different $\theta$ are shown in Fig.\ref{fig difftheta}. We are looking for an optimal number of $\theta$, using which can provide us good cross correlation between DM and $n_{gal}$. We can see that considering different redshift, $\theta=0.0027$ is an optimal choice and for simplicity we choose half top-angle for both 'cone' and 'cone+' to be $\theta= 0.003$ (radiant) which is approximately equal to 1' in degree. 
The mollweide projection images are shown in the Fig.\ref{fig mollview}. At low redshift, we will encounter many pixels whose value is 0, in the mollview image of cone. Such a problem can be solved by using 'cone+'. Hence the cross-correlation between DM and $n_{gal}$ will be better if we use 'cone+' at low redshift. But at high redshift, the little difference between 'cone' and 'cone+' will not make significant influence on the cross-correlation. The comparison is shown in Fig.\ref{fig conevscone+}. We have compared the cross correlation coefficient between DM and $n_{gal}$ using 'cone' and 'cone+', and we have also shown the redshift evolution of the cross correlation using 'cone'. It is clear that at higher redshift, the correlation is better and the difference between 'cone' and 'cone+' is smaller. Because 'cone+' require more complicated calculation for the galaxy distribution and considering that we are more interested in high redshift FRBs and their DM,  we will use the 'cone' method to do further analysis about DM.

\section{\label{sec 5}MCMC constrain}
\begin{figure}
    \centering
    \includegraphics[width=1\columnwidth]{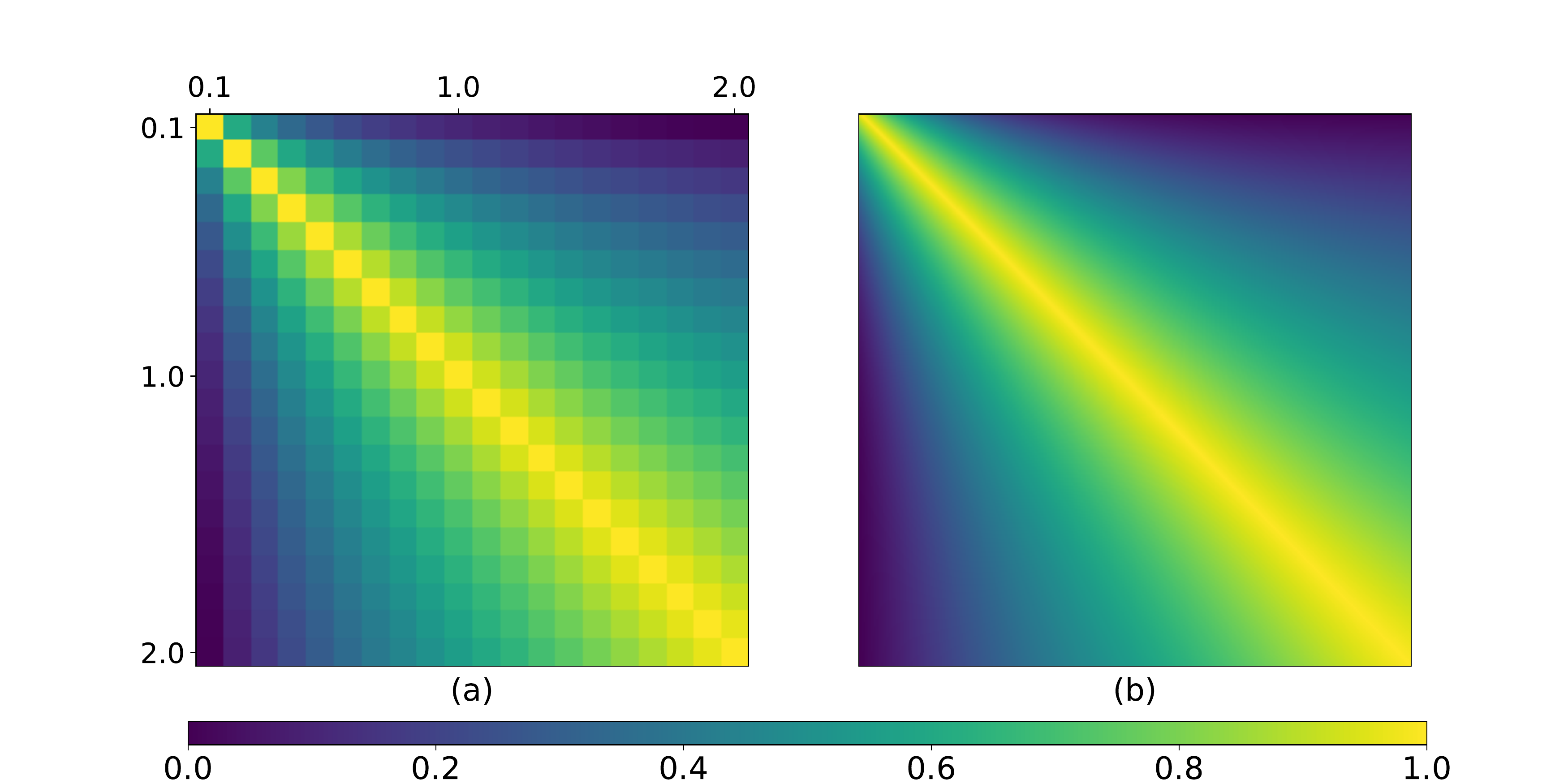}
    \caption{The $20 \times 20$ matrix of Pearson product-moment correlation coefficients obtained from the data is shown on the left, and the right one is the matrix we obtained from fitting function, which is smooth.}
    \label{fig covmn}
\end{figure}

\begin{figure}
    \centering
    \includegraphics[width=1\columnwidth]{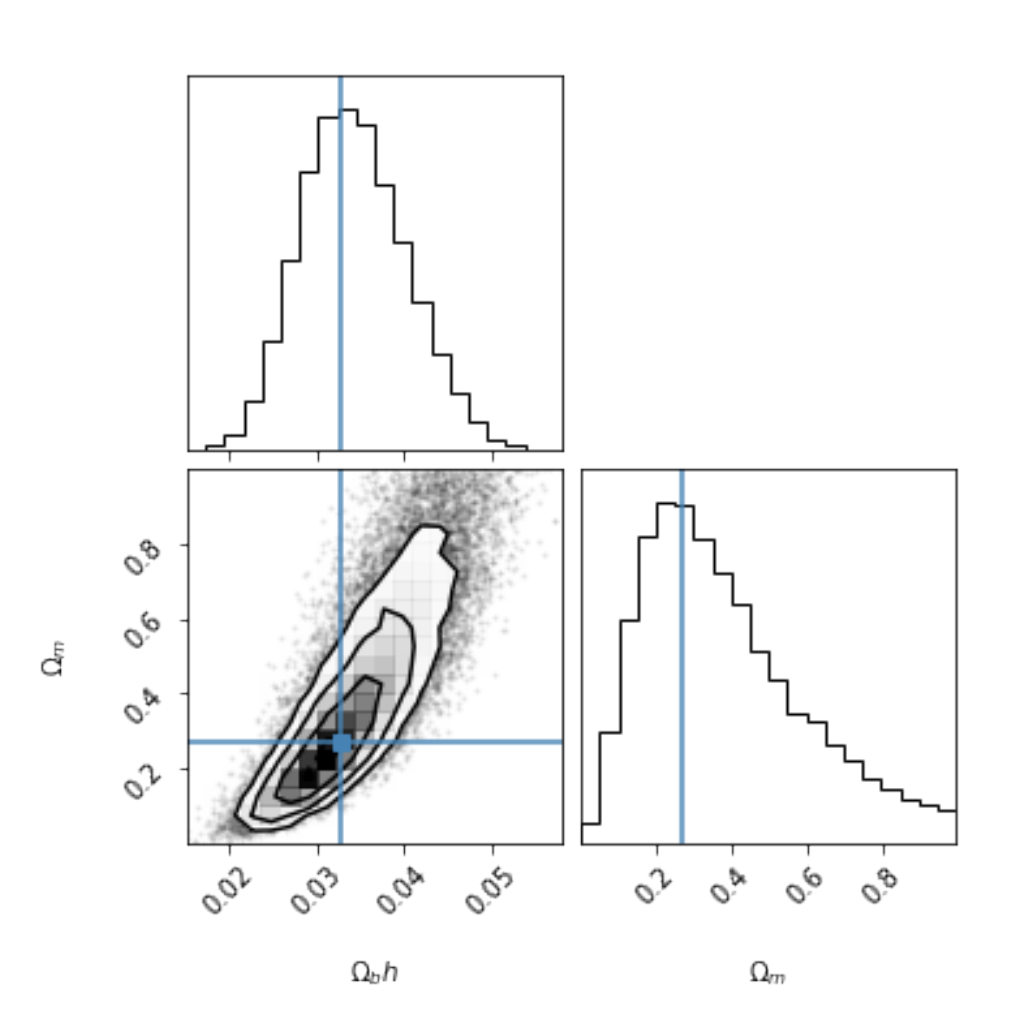}
    \caption{Here we show the MCMC result with using 500 FRBs without correction. The blue line marks the true value of parameters. The ture value lies in the 1-$\sigma$ area.}
    \label{fig MCMCtrue}
\end{figure}

\begin{figure}
    \centering
    \includegraphics[width=1\columnwidth]{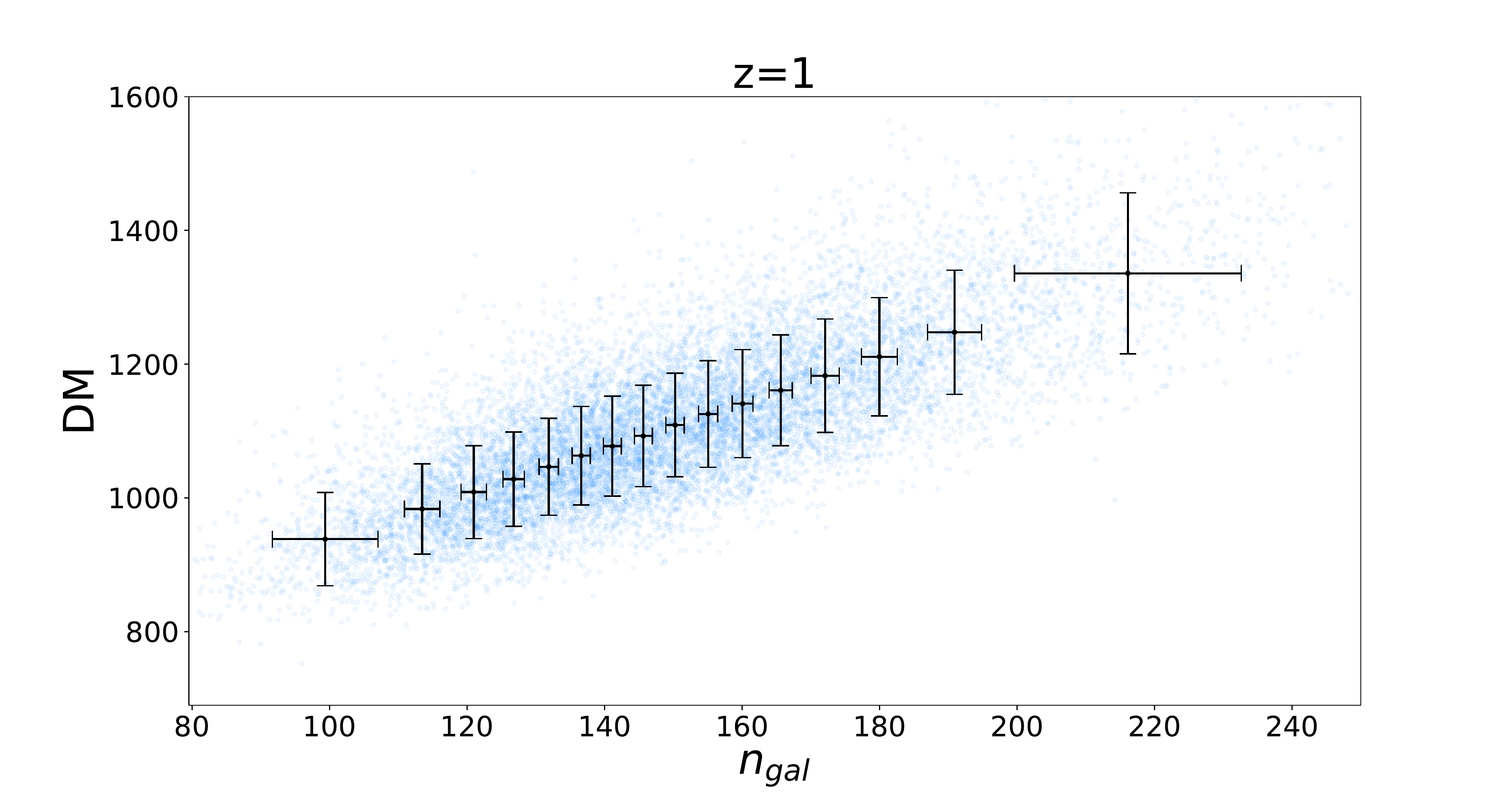}
    \caption{The linear relation between DM and $n_{gal}$ at z=1. We show the 16 groups' average value with their standard deviation in both direction. And the original values are decorated as translucent blue dots in the background.}
    \label{fig linear}
\end{figure}

We will follow the formula Eq.\ref{eq DMIGM} to do MCMC analysis with several FRBs randomly obtained in the simulation. In our research, we will not assume any prior knowledge about the parameters; hence all the parameters will follow uniform distribution in their well-defined interval ($\Omega_b,\Omega_m \sim u(0,1)$). We randomly set points in the simulation box to be FRBs source and use the same method to calculate their DM. Then we will try to find a method to associate the DM with $n_{gal}$, and finally compare the results. 

We have obtained a $20 \times 20$ covariance matrix from the 20 mocks built from the simulations at 20 redshifts, to acquire the covariance matrix of FRBs at arbitrary redshift, We use them to fit a function of the form 
\begin{equation}
    R(z_i,z_j)=b(z)e^{-\frac{\Delta z}{a(z)}}+(1-b(z))
    \label{eq covmterm}
\end{equation}
where $R(z_i,z_j)$ is the $(i-j)^{th}$ term of Pearson product-moment correlation coefficients; $\Delta z=z_j-z_i, (z_j>z_i)$ is their redshift difference; $a(z), b(z)$ are our fitting parameters. We then use a third-order polynomial $p(z)=p_3\times z^3 +p_2\times z^2 + p_1\times z +p_0$ to fit a(z) and b(z), the results are shown in table \ref{tab covm}. In Fig.\ref{fig covmn}, We also plot a smoother image of Pearson product-moment coefficient matrix compared to original one with our fitting function, in order to verify whether it is appropriate. The result shows it is applicable within the range of our research. The MCMC result from 500 FRBs with the location of true value is shown in the Fig.\ref{fig MCMCtrue} The true values well lie in the $1-\sigma$ area, so that we use the covariance matrix and the related method to do further study.
\begin{table}
    \centering
    \caption{Fit Parameters}
    \begin{ruledtabular}
    \begin{tabular}{c|cccc}
        & $p_0$ & $p_1$ & $p_2$ & $p_3$  \\\hline
        a(z) & 0.122 & 1.505 & -1.300 & 0.500\\
        b(z) & 0.857 & -0.494 & 0.192 & -0.014
    \end{tabular}
    \end{ruledtabular}
    \label{tab covm}
\end{table}

From Eq.\ref{eq DMIGM}, we can see that the cosmological parameters that we can put constraints using DM(z) is $\Omega_M$, $\Omega_b$ and $H_0$, under the flat space assumption. However $\Omega_b$ and $H_0$ are degenerate, the only independent cosmological parameters are $\Omega_M$ and $\Omega_b h$. Therefore, it is sufficient to perform MCMC fitting for these two sets of parameters. In Fig.\ref{fig MCMCtrue}, we have shown that using the covariance matrix fitting function and the randomly generated FRBs, we can put constraints on the cosmological parameters and the true value well lie in the $1-\sigma$ range. In the error range contour, it is well known that the large scale structure plays an important role in the error budget of the function DM(z). With the information of line-of-sight galaxy number, we can have a better estimation of the large scale structure of matter distribution, thus provide a smaller error.

\subsection{Correction Method}
\begin{figure}
    \centering
    \includegraphics[width=1\columnwidth]{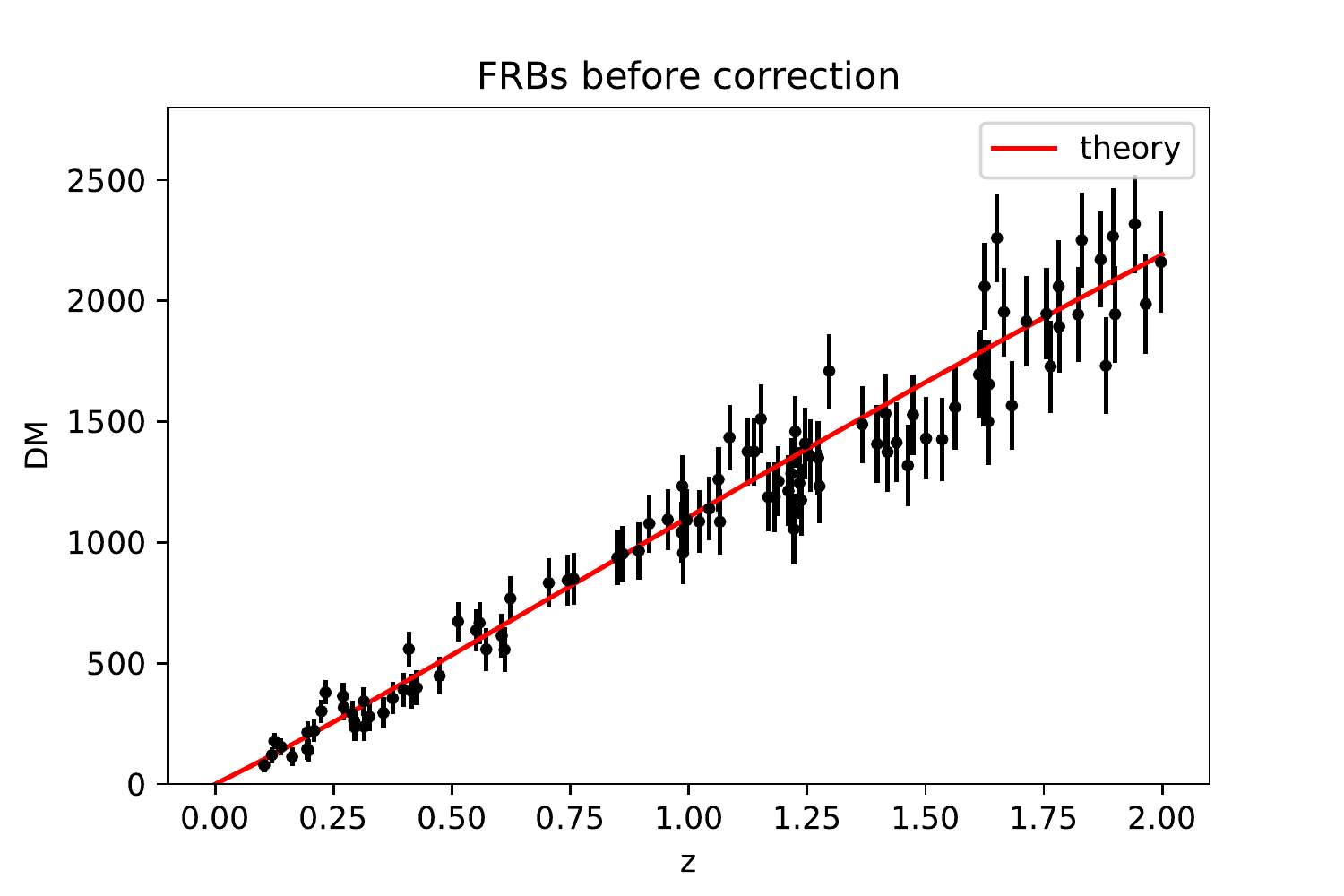}
    \includegraphics[width=1\columnwidth]{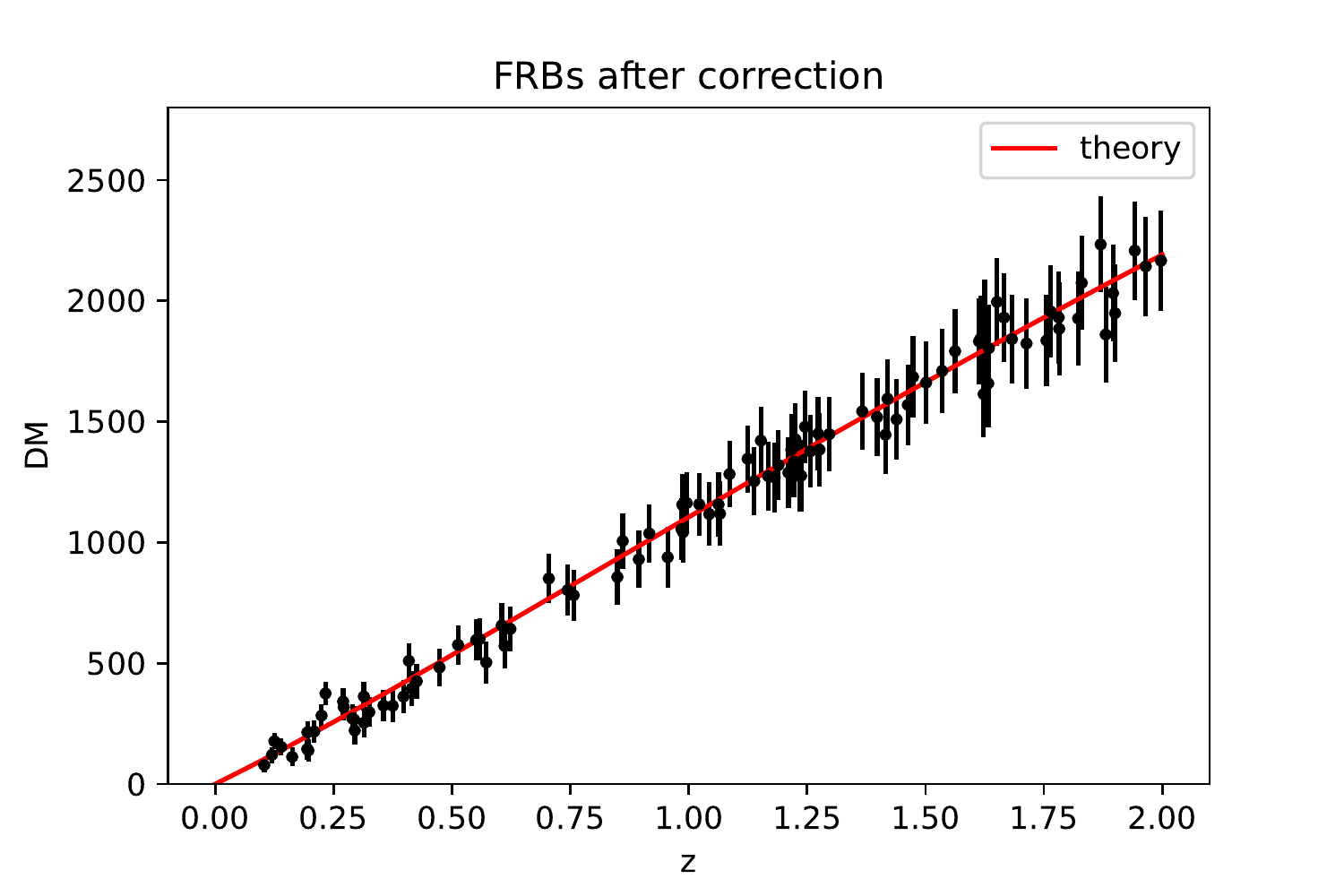}
    \caption{On the top, we show the original DM distribution of simulated FRBs, together with the theoretical prediction line. On the bottom, we show the corrected DM distribution of simulated FRBs using the line-of-sight galaxy number information, together with the theoretical prediction line. It is clear that with the galaxy number information provided, the DM deviate less from the theoretial curve with correction.}
    \label{fig: dmzcurve}
\end{figure}
According to the value of $n_{gal}$, we divided the 196608 pixels into 16 groups equally and calculate the standard deviation and the average of each group. Their distribution is shown in Fif.\ref{fig linear}. Then we figure out that the relation between $n_{gal}$ and DM is almost linear when $z \geq 0.3$; but if z is too small, there will be too many pixels with its $n_{gal}$ is 0. With a determined halo occupation distribution model, the mean $n_{gal}$ is known. We use these 16 groups of points to fit a function of form separately at different redshift,
\begin{equation}
    DM(n_{gal},z)=k(z)(n_{gal}-\bar{n}_{gal})+b(z)
    \label{eq linear}
\end{equation}
As we expect, b(z) is equal to $DM_{mean}(z)$ within 4\% error, and the error decreases to 0.4\% as z increases to 2.

Strictly speaking, k(z) is not only dependant on the $\theta$ of the light-cone but also related to the halo occupation model we established. Hence here we will only use interpolation method to obtain k(z) at FRBs' redshift, and leave further research for our future study. 

With known $n_{gal}$ and DM associated with the same FRB which located at $z>0.2$, we can correct our expectation about corresponding $DM_{mean}$ by using Eq.\ref{eq covmterm},
\begin{equation}
    DM_{cor}=DM_{frb}-k(z)[n_{gal}-\bar{n}_{gal}],
    \label{eq correc}
\end{equation}
where $DM_{frb}$ is the primitive DM, and $DM_{cor}$ is what we will use in MCMC analysis after correction.

\subsection{Result Improvement}

\begin{figure}
    \centering
    \includegraphics[width=1\columnwidth]{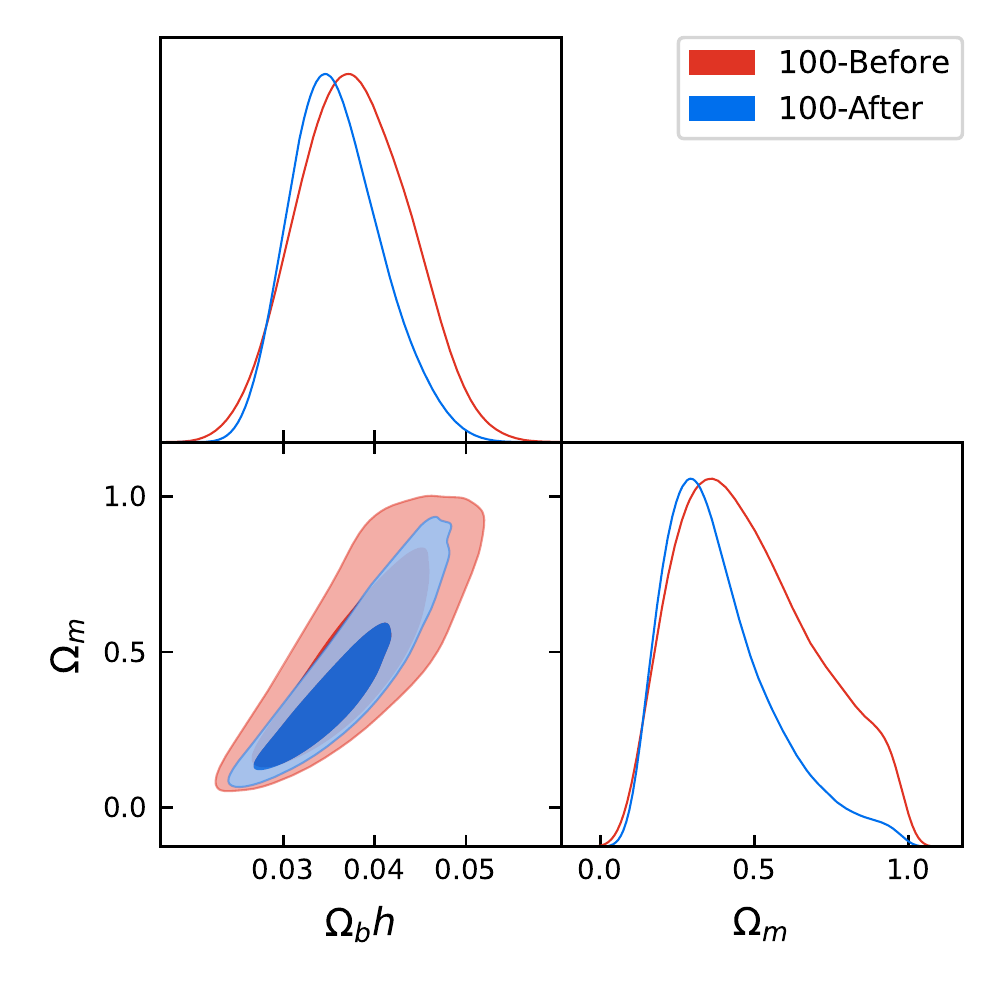}
    \includegraphics[width=1\columnwidth]{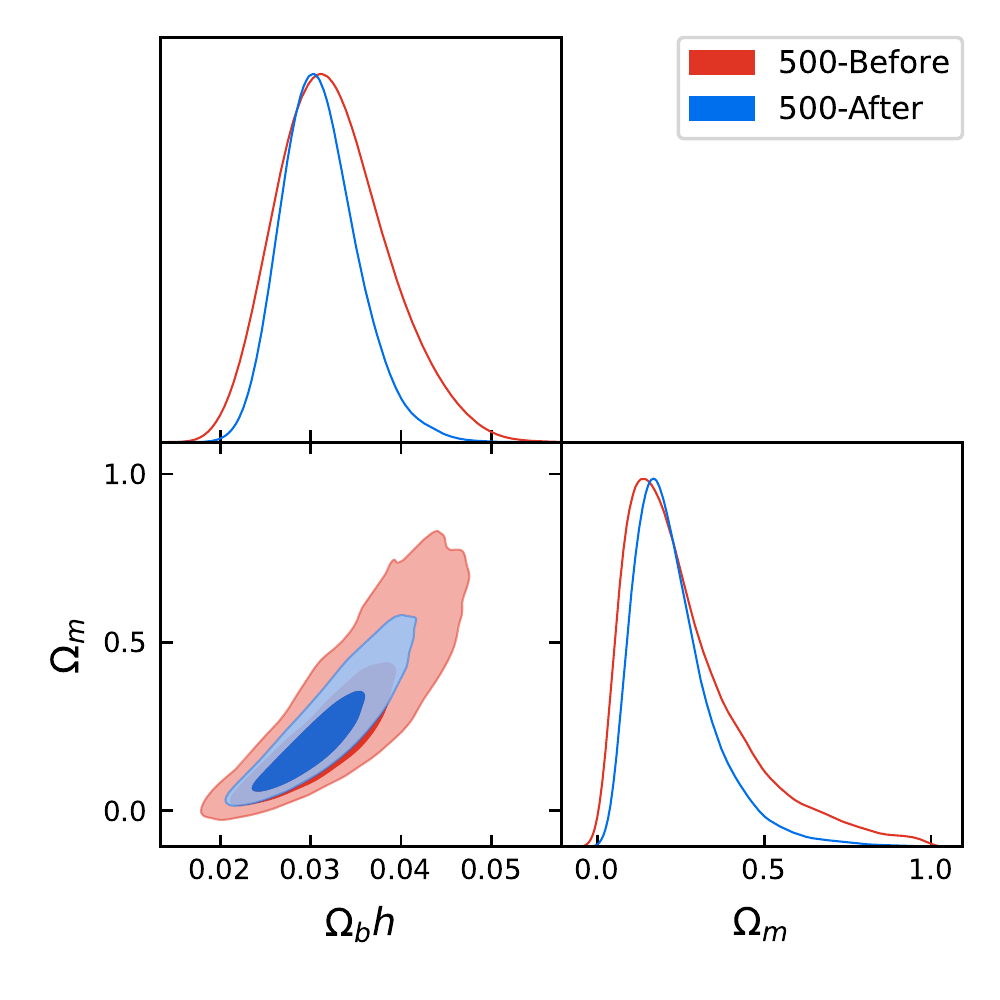}
    \caption{On the top, we show the improvement of cosmological constraints using 100 FRBs. On the bottom, we show the improvement using 500 FRBs. We can see clear improvement in the cosmological parameters constraints for both cases.}
    \label{fig: mcmcimprove}
\end{figure}
\begin{table}
    \centering
    \begin{ruledtabular}
    \caption{Cosmological Parameter Constrains}
    \footnote{we quote the uncertainties based on the 16th, 50th, and 84th percentiles of the samples, and in our case, it is a little different from the maximum likelihood point.}
    \begin{tabular}{c|cc}
     & $10^2 \Omega_b h$ & $10 \Omega_m$ \\\hline
    True & 3.283 & 2.7 \\
    &&\\
    100-before & $3.7606^{+0.6367}_{-0.6004}$ & $4.58^{+2.79}_{-2.01}$\\
    &&\\
    100-after & $3.5468^{+0.5506}_{-0.4570}$ &$3.52^{+2.15}_{-1.36}$\\
    &&\\
    500-before & $3.2030^{+0.6572}_{-0.5537}$&$2.19^{+2.19}_{-1.22}$\\
    &&\\
    500-after &  $3.0776^{+0.4515}_{-0.3834}$& $2.06^{+1.36}_{-0.89}$
    \end{tabular}
    \label{tab result}
    \end{ruledtabular}
\end{table}
With the information of line-of-sight galaxy number provided, it is expected that we can have a better estimation of DM, partially get rid of the error introduced by large scale structure. As the results shown in Fig.\ref{fig: dmzcurve} suggest, our correction is effective, the DM of FRBs distributes closer to the theoretical expectation compared with original data. 
More practically, if we apply our correction method to put constaints on the cosmological parameters given 100 random FRBs or 500 random FRBs, the improvement of constraints can be clearly seen in Fig.\ref{fig: mcmcimprove}. The blue contours show the parameter uncertainty range after the correction and the red contours show the range before the correction, which is what we normally see in literature. The detail number of the cosmological parameters are summarised in table \ref{tab result}. We shall notice that such improvement is free. Since if we can localize the FRBs and get their redshift, these FRBs are localized by matching the radio signal location to the given galaxy catalog. We naturally know the number of galaxies near the host galaxies of the FRBs, which can be easily obtained. The improvement of the precision of cosmological constraints is more than 20\%.

\section{\label{sec 6}discussion}
With more and more FRBs being discovered and localized, FRB cosmology has attracted quite some attention in the cosmology study. Using dispersion measure to measure the expansion history of the universe is a unique method. If we have better understanding of the error budget in the dispersion measure of FRBs, we will have better chance to put tight constraints on cosmological parameters. 

In this paper, we have proven the idea that, using the information of large scale structure traced by galaxies, we may put tighter constraints with FRBs, really works. We have shown that just use the galaxy number within 1' around the FRBs host galaxies, we can correct the DM(z) expectation value optimally. Afterwards, we may have more than 20\% improvement in the cosmological constraining precision.

We have covered almost all the steps for cosmological analysis using FRBs in this paper, but there are quite some details need to be improved so that our method can be really applied to the real observation. For example, how do the DM-$n_{gal}$ relation rely on the galaxy mean number density, what is the parameters of this relation for some real observational data, is the simple model of ionized gas sufficient for our analysis here, how much improvement will we have if we consider other error budgets etc. We will cover these questions in our future study. 
Beyond cosmological application, the cross correlation between DM and galaxy distribution can also tell us more about the gas environment around galaxies and the properties of IGM. We hope our study here can provide useful hints for the other study in the future.

\begin{acknowledgments}
J.Z acknowledges support from the Ministry of Science and Technology of China (grant Nos. 2020SKA0110102). We thank Chuhan Jiang, Yao Zhang and Fayin Wang for useful discussion.
The CosmoSim database used in this paper is a service by the Leibniz-Institute for Astrophysics Potsdam (AIP).
The MultiDark database was developed in cooperation with the Spanish MultiDark Consolider Project CSD2009-00064.
We use the MDR1\cite{riebe2013multidark} database with its FOF\cite{prada2012halo} halo catalogue as the simulated universe. We use python packages \textbf{emcee}\cite{foreman2013emcee} for MCMC analysis; use \textbf{getdist}\footnote{\url{https://github.com/cmbant/getdist}} and \textbf{corner}\footnote{\url{https://github.com/dfm/corner.py}} for plotting.
\end{acknowledgments}

\bibliographystyle{apsrev}
\bibliography{Amyref}% Produces the bibliography via BibTeX.

\end{document}